\documentclass[titlepage,12pt]{article}
\usepackage{amssymb}
\usepackage{amsfonts}
\begin{document}

{\Large \bf Towards bringing Quantum Mechanics and \\ \\ General Relativity together} \\ \\

{\bf Elem\'{e}r E Rosinger} \\
Department of Mathematics \\
and Applied Mathematics \\
University of Pretoria \\
Pretoria \\
0002 South Africa \\
eerosinger@hotmail.com \\ \\

{\bf Abstract} \\

Two questions are suggested as having priority when trying to bring together Quantum Mechanics
and General Relativity. Both questions have a scope which goes well beyond Physics, and in
particular Quantum Mechanics and General Relativity. \\ \\

{\large \bf 1. Preliminary Remarks} \\

Bringing Quantum Mechanics and General Relativity together is nowadays considered to be a
problem of Physics, and in fact, its fundamental one. Yet not seldom when solving more
fundamental problems in any given realm, it may well happen that the sought after solution is
less accessible when the ways of thinking are constrained or limited to the respective realm,
in this case, the usual ones in Physics. \\

Let us recall in this regard the celebrated 1960 paper of the Nobel laureate Eugene Wigner,
entitled "The unreasonable effectiveness of mathematics in the natural sciences", see [2].
There have been various comments upon, and interpretations of that quite manifest and most
impressive "unreasonable effectiveness". And the issue, most likely, is not that Mathematics
happens to be more fundamental than Physics, for instance. Rather, it could be about the fact
that both Mathematics and Physics, as they stand today, come from a yet deeper mode of human
insight, a mode which, so far, has not been formalized as a science. \\
In view of such a possibility, it may indeed be useful to try to avoid the exclusiveness of
constraining the ways of thinking to the customary ones in Physics, when trying to bring about
the unification of Quantum Mechanics and General Relativity. \\

The ways of thinking in both Mathematics and Physics have their specific qualities and
strengths. In Mathematics one is supposed to be perfectly precise, and also most abstract, at
least when compared with other natural sciences. Also, creativity in bringing up new ideas and
results is rigorously controlled by the requirements of logic, as well as of proofs of
conjectures. On the other hand, Physics is significantly more free and protean when it comes
to new ideas and hypotheses. Of course, there is again a control, and in fact, a double one
this time, namely, of a certain theoretical consistency and of possible supporting
experiments, or at least, the lack of contrary ones. The theoretical consistency, however, is
not of that extreme rigor as in Mathematics, since much of the more fundamental Physics, among
others Quantum Mechanics or Quantum Field Theory, has a somewhat tentative or heuristic
mathematical formulation. \\

This specific quite free and protean nature of thinking in Physics, rather distinct from that
in Mathematics among others, brings with it a certain entrapping temptation. Namely, it
creates the strong and often irresistible impression, if not in fact of certainly, not only of
a generously rich self-contained realm of thinking, but implicitly also of the
inappropriateness of any other way of thinking when dealing with problems in Physics. \\

So much, therefore, for the chance to realize that more fundamental problems in Physics may
seriously benefit from ways of thinking which are not constrained or limited to the usual ones
in Physics. \\

And now, let us turn to some of the specifics of the problem of bringing Quantum Mechanics and
General Relativity together. \\

In this regard, it may be instructive to start by recalling the way Special Relativity seems
to have emerged in Einstein's thinking. It is often reported that, around the age of 16 or 17,
Einstein started to wonder about the following thought-experiment. Assume that a beam of light
is emitted from some source S and in the direction A. Einstein then imagined that he himself
would move next to that beam of light and do so with the velocity of light. The question which
kept puzzled him was : what would he observe ? His answer was that he would only observe some
stationary states. For instance, he would keep seeing the source S of light in ever the same
state in which it was when it emitted that beam. Further, he would see the light beam next to
him as a standing wave in space. And obviously, both of these observations where wrong. After
all, the source may change its state with the passing of time, not to mention that standing
light waves were not compatible with the Maxwell equations. \\
The conclusion Einstein drew in 1905 from the above was that the velocity of light could not
depend on that of the observer. \\
This conclusion then became one of the two basic principles of Special Relativity. The other
basic principle was the old Galilean one, according to which it is not possible to detect
absolute rest or absolute motion. \\

What happened during the next ten years, while Einstein tried to include gravitation in
relativity, may be particularly instructive today, we one attempts to bring together Quantum
Mechanics and General Relativity. \\
The various respective attempts Einstein made over a decade were motivated not so much by what
would later be called "general covariance", as rather by a number of specific physical
arguments. On the other hand, focusing more on "general covariance" may have helped in
simplifying the issues and reaching in a more direct manner the Einstein field equations.
After all, there are only two entities invariant under "general covariance", namely, volume
and curvature. And the true essence of General Relativity is very much contained in the
"general covariance" of the Einstein field equations. \\

A similar situation was to happen half a century later in Quantum Mechanics with respect to
the Bell Inequalities. As it turns out, Bell type inequalities were known to George Boole in
the 1850s, see [1]. Consequently, the Bell Inequalities can be established without any
physical arguments. It follows that the original contribution in the Bell inequalities is not
in the inequalities themselves, but in showing the fact that they contradict Quantum
Mechanics. On the other hand, the clarity of this most simple and fundamental fact is
completely lost in the way the Bell inequalities are rather without exception presented and
commented upon in a considerable amount of texts written by physicists. Indeed, in all such
texts, which limit
themselves exclusively to the exhibition of any number of arguments in Physics, the resulting
complex buildup of arguments can only serve to obscure the underlying clarity and simplicity
of the mentioned fundamental fact. \\

{\large \bf 2. What Is the Local Point of View ?} \\

There appear to be two alternatives when formulating theories of Physics, namely

\begin{itemize}

\item (A 1) There can only be a non-local, more precisely, global formulation, which therefore
must be unique, or

\item (A 2) A variety of local observers can have valid formulations regarding the same
physical wholeness which is the object of the theory, just as it is the object in the case of
the first alternative (A 1).

\end{itemize}

So far, throughout the history of Physics, the second alternative (A 2) has been embraced.
Moreover, there has been a significant unease, if not even ill-feeling, with respect to
physical phenomena which cannot be localized in some suitable manner. \\

Let us therefore start from the point of view of the second alternative (A 2). \\

In this case, each local formulation of the physical wholeness is expected to be equivalent in
some appropriate way with all the other ones. In other words, we expect a certain principle of
"relativity" to hold among the set of such local formulations. \\
For instance, in Special and General Relativity this is precisely the case. Furthermore, in
these two theories the local aspect corresponds to a given frame of reference, while the
equivalence is expressed by the Lorentz, respectively, general covariance. \\

Turning now to the bringing together of Quantum Mechanics and General Relativity, and doing so
along the lines of the second above alternative (A 2), that is, by building local theories,
the first question which appears to arise is as follows

\begin{itemize}

\item (Q 1) : What is the appropriate local point of view ?

\end{itemize}

Clearly, the scope of this question is more general than referring only to Quantum Mechanics
or General Relativity. In fact, the scope of this question may easily go beyond the whole of
Physics as such. Yet its importance, and in fact, priority is quite obvious, in case we start
from the point of view of the second alternative (A 2). \\

Now, coming again back to our problem of bringing Quantum Mechanics and General Relativity
together, it is not immediate that the local frames of reference, so essential in both Special
and General Relativity, may be sufficient in this case. Also, the non-relativistic classical,
or special relativistic frames of reference presently used in Quantum Mechanics may equally be
insufficient. \\

And then, before unleashing the usual protean variety of exclusively physical arguments, it
may perhaps be more appropriate to focus on this seemingly more general and deep issue,
namely, to try to understand what may an appropriate local point of view be, as asked in (Q
1). \\

Here, in this regard, it may be useful to recall the following. \\

In General Relativity there exists a "state space" given by a specific four dimensional
Einstein manifold.  In non-relativistic Quantum Mechanics of finite systems there exists a
"configuration space" given by a suitable Euclidean space, while the "state space" is supposed
to be the Hilbert space of square integrable functions defined on the Euclidean space. \\
This difference alone, not to mention that related to the way the spaces of "observables" are
defined in each of these two theories, may already be reason enough to consider the above
question (Q 1). \\

What has instead happened so far is the following. An exclusive focus was placed on coming up
with suitable new "state" or "configuration" spaces. String and Super-string Theory, for
instance, postulate as ground physical entities certain continuous geometric structures which
take place of the zero dimensional points of classical "state spaces". Alternative approaches,
such as in Quantum Gravity, may postulate as an ultimate underlying physical ground various
discrete, or locally discrete structures. \\

Thus none of these approaches sees as a priority answering the above question (Q 1).
Consequently, none of these approaches can express in a suitable manner "general covariance",
which is the hallmark of the second, that is, local alternative mentioned above in (A 2). \\

It appears, therefore, that the priority is indeed with answering the above question (Q 1), or
in its reformulated manner

\begin{itemize}

\item (Q 1$^*$) What is the appropriate concept of "frame of \\ reference" ?

\end{itemize}

\bigskip

{\large \bf 3. Which General Covariance ?} \\

Once question (Q 1), or equivalently, (Q 1$^*$) was answered, one is led to the second
question

\begin{itemize}

\item (Q 2) What are the requirements of "general covariance" ?

\end{itemize}

In this regard, it may be quite likely that an answer to question (Q 1) does not necessarily
determine, but only limits or conditions the answer to question (Q 2). \\

{\large \bf 4. Conclusions} \\

Needless to say, lots of physical arguments may be involved in answering the above questions
(Q 1) and (Q 2). \\

However, two facts should not be overlooked, namely

\begin{itemize}

\item Both questions (Q 1) and (Q 2) have a scope which goes well beyond Physics, and in
particular Quantum Mechanics or General Relativity,

\item Both questions (Q 1) and (Q 2), in that order, seem to have the highest priority when
trying to bring together Quantum Mechanics and General Relativity.

\end{itemize}

Within alternative (A 2) - which is both the traditional and present day view of Physics -
until the emergence of Special and General Relativity it appeared most natural to start first
with a view of the physical wholeness which is the object of any given theory in Physics. This
wholeness was then modelled, among others, by one or another "state space". Yet ironically, no
such "state space" could be defined mathematically, unless some frame of reference was tacitly
assumed and employed. Typical in this regard is the absolute space and time of Newtonian
Mechanics. In this way, the priority of "frames of reference" was in fact already there, even
if implicitly. \\
Starting with Special Relativity, "frames of reference" obtained an explicit priority in any
mathematical model. \\
A second departure happened with Quantum Mechanics, where in addition to the traditional
"state space" given by a suitable Hilbert space, one would now have the "observables" given by
self-adjoint operators on that Hilbert space. And strangely enough, the "state space" would
now no longer be assumed directly accessible, but only through the "observables". \\

Such a state of affairs may further support the above point of view regarding the priority in
asking questions (Q 1) and (Q 2). \\

\end{document}